# The Current State of AI Bias Bounties: An Overview of Existing Programmes and Research

Sergej Kucenko[1], Nathaniel S. Dennler[2], Fengxiang He[3]

1. Department of Health Sciences, Hamburg University of Applied Sciences, Hamburg, Germany, sergej.kucenko@reuben.ox.ac.uk
2. Computer Science & Artificial Intelligence Laboratory, Massachusetts Institute of Technology, Cambridge, United States, dennler@mit.edu
3. School of Informatics, University of Edinburgh, Edinburgh, Scotland, F.He@ed.ac.uk

2 October, 2025

**Abstract**

Current bias evaluation methods rarely engage with communities impacted by AI systems. Inspired by bug bounties, bias bounties have been proposed as a reward-based method that involves communities in AI bias detection by asking users of AI systems to report biases they encounter when interacting with such systems. In the absence of a state-of-the-art review, this survey aimed to identify and analyse existing AI bias bounty programmes and to present academic literature on bias bounties. Google, Google Scholar, PhilPapers, and IEEE Xplore were searched, and five bias bounty programmes, as well as five research publications, were identified. All bias bounties were organised by U.S.-based organisations as time-limited contests, with public participation in four programmes and prize pools ranging from $7,000 to $24,000. The five research publications included a report on the application of bug bounties to algorithmic harms, an article addressing Twitter's bias bounty, a proposal for bias bounties as an institutional mechanism to increase AI scrutiny, a workshop discussing bias bounties from queer perspectives, and an algorithmic framework for bias bounties. We argue that reducing the technical requirements to enter bounty programmes is important to include those without coding experience. Given the limited adoption of bias bounties, future efforts should explore the transferability of the best practices from bug bounties and examine how such programmes can be designed to be sensitive to underrepresented groups while lowering adoption barriers for organisations.

## 1. Introduction

Artificial intelligence (AI) is becoming increasingly present in our lives and has the potential to transform various industries and sectors, such as healthcare [1], finance [2], and education [3]. However, the rapid advancement of AI technologies also brings risks to society. One such risk is bias, which can be defined as 'a systematic error in decision-making processes that results in unfair outcomes' [4]. AI bias can stem from three different sources: *data bias*, characterised by incomplete or unrepresentative training data, where certain groups or characteristics are over- or underrepresented; *algorithmic bias*, characterised by algorithmic design choices that embed biased assumptions or decision criteria and result in differential outputs; and *user bias,* characterised by individuals introducing conscious or unconscious biases through the provision of biased training data or through interactions with the AI system [4].

Numerous reported cases of harmful AI bias have raised serious concerns about the potential negative impacts of AI. For instance, Amazon abandoned an AI recruiting tool after the tool was demonstrated to be biased against women. The tool was trained on the company's historical resume data, which was skewed toward male applicants. As a result, it favoured resumes that included male-dominated terms and penalised those with terms associated with women [5]. Another example is an algorithm used by the municipality of Rotterdam to predict social welfare fraud, which resulted in discriminatory fraud predictions. This AI bias manifested as particularly young single mothers who had limited Dutch language skills were subjected to fraud investigations based on biased training data, with some having their social benefits wrongfully lowered [6].

Various approaches to identify biases in AI have been proposed [7-8]. However, the methods employed by many companies, including bias evaluation benchmarks and dataset and model documentation, rarely necessitate a company to engage with the communities impacted by AI harms [9].

One method that directly involves these communities is bias bounties [9]. AI bias bounties build on bug bounty programmes. In bug bounties, individuals external to the bounty-hosting organisation are rewarded for identifying security vulnerabilities in software. Similarly, bias bounties are crowdsourced initiatives in which organisations incentivise participants to detect and report biases in AI systems, aiming to improve system fairness and prevent harms. The promise of bias bounties lies in increasing scrutiny of AI systems by motivating those unaffiliated with the AI-developing organisation to report bias through financial incentives in a formal programme [10]. They also have the potential to introduce a broader range of perspectives into bias detection and to promote accountability. The benefits for the bias bounty-hosting organisation include mitigation of AI bias and the associated risks (e.g., legal issues, reputational damage), along with potential reputational gains as a 'socially responsible' organisation.

The idea of offering bounty programmes for AI bias was suggested in two different articles in 2018 [11-12]. This method has gained recognition since then, with recommendations to conduct bias bounties made in the UK Government's *Ethics, Transparency and Accountability Framework for Automated Decision-Making* [13] and by the National Institute of Standards and Technology in the United States [14]. However, to the best of our knowledge, no work has yet presented the current landscape of AI bias bounties. Such a survey would be valuable for informing the design of future AI bias bounties and research in this area by identifying existing practices and highlighting current gaps and challenges.

Therefore, this survey aims to 1) identify and analyse existing AI bias bounty programmes and 2) present available research publications on bias bounties. It is intended to support both organisations planning to conduct bias bounty programmes and researchers by synthesising existing efforts and offering practical recommendations.

## 2. Methods

Given that scientific articles are unlikely to be the primary source of information on AI bias bounties by organisations, the search for such programmes was conducted using the Google search engine. Academic publications related to AI bias bounties were searched for in Google Scholar, IEEE Xplore, and PhilPapers. Both existing programmes and research publications were searched using the following sets of keywords: 1) artificial intelligence, AI, machine learning, ML, algorithm; 2) bias; and 3) bounty or bounties. The query strings for each database and the number of search results are provided in the appendix. Records published between 2017 and 2024 were searched during the week of November 18 to 24, 2024. This time frame was chosen due to the novelty of bias bounties and the rapidly evolving nature of the AI field. A programme was considered a bias bounty if it offered a monetary reward to contributors external to the hosting organisation for identifying and reporting AI-related biases. Only records that presented such an AI bias bounty programme or research on bias bounties were included. Records not published in English or not available through open access or institutional log-in were excluded. The search for records, as well as their inclusion and exclusion, was conducted by one researcher (SK). Depending on the potential relevance of a record, exclusion decisions were based on the title alone, the title and abstract, or, where applicable, the full text.

In addition to describing each AI bias bounty programme and research publication in the Results section, we analysed the bias bounties using design levers proposed by Kenway et al. (Figure 1) in their report on applying bug bounties to algorithmic harms [15]. These design levers, which differentiate bug bounty programme design through various aspects, were previously applied to Twitter's bias bounty challenge in their work [15]. We adopted this framing to highlight commonalities and differences among the bias bounties presented in this survey. Only publicly available information was used for the analysis.

| TARGET ENTITIES | Voluntary | | Adversarial | |
|---|---|---|---|---|
| | Only reports relating to organizations that have consented to receiving such reports are accepted. | | Reports related to organizations that have not agreed to receive such reports are also accepted. | |
| COMPENSATION MODEL | Non-Monetary | Bounties | Contract | Employment |
| | Security researchers receive only non-monetary benefits in exchange for their findings. | Organizations or platforms pay security researchers for finding in-scope vulnerabilities, with rediscovery of already-identified vulnerabilities genverally not rewarded. | Organizations or platforms retain security researchers on a temporary, contractual basis to undertake specific services. Researchers are compensated regardless of findings. | Organizations or platforms retain security researchers on a permanent basis to undertake wide-ranging research. Researchers are salaried and receive typical employment benefits. |
| DISCLOSURE MODEL | Delayed Full Disclosure | Coordinated Disclosure | | Non-Disclosure |
| | Researchers can freely disclose their findings to the public on a predetermined time frame without additional approval from the affected organization. | Organizations contract with a BBP platform to provide specific services within or related to their BBP, while handling other elements in-house. | | Researchers cannot publicly disclose their findings. |
| PARTICIPATION MODEL | Public | | Private / Invite-Only | |
| | All researchers are invited to conduct research and submit reports. | | Only pre-authorized researchers are invited to conduct research and submit reports. | |
| PROGRAM MANAGEMENT | Platform-Managed | Mixed Management | | Self-Managed |
| | Organizations contract with a BBP platform to deliver their BBP, with reports typically channeled through a specific portal on the platform. The platform also provides related services (e.g., report validation, triage, patch verification, etc.). | Organizations contract with a BBP platform to provide specific services within or related to their BBP, while handling other elements in-house. | | Organizations handle delivery of their BBP in-house. They accept reports directly through their websites or via a dedicated email address and handle validation, triage, andpatch verification internally . |
| PROGRAM DURATION | Ongoing | | Time-Limited | |
| | Reports are accepted on a continuous basis for an evolving range of assets. | | Reports are accepted for a specified range of assets over a limited time period. | |
| PROGRAM SCOPE & ACCESS | Constrained ⟷ | | Expansive | |
| | Only a limited variety of vulnerability types and / or systems are identified as in-scope of the program. | | All possible vulnerability types and systems are in-scope of the program. | |
| | 'Closed Box' ⟷ | | 'Open Box' | |
| | Testing is limited to publicly available resources and tooling, without additional organizational or technical access (e.g., to documentation or source code). | | Additional access, either organizational or technical, is provided to enable a deeper level of testing. | |

Figure 1: Algorithmic Justice League's design levers for bug bounty programmes [15]. Reproduced with permission from the Algorithmic Justice League (https://www.ajl.org/bugs).

## 3. Results

The AI bias bounty programmes and research cited in this section were identified through a Google search (216 results) and a Google Scholar search (first 100 results screened). Searches in IEEE Xplore and PhilPapers using the strategy presented in the appendix did not yield any results.

### 3.1. Existing AI Bias Bounty Programmes

Five AI bias bounties, all conducted by U.S.-based organisations, were identified (Table 1) [16-20]. Twitter introduced the first bias bounty in 2021 [16]. Subsequently, initiatives were launched by the nonprofit organisations Bias Buccaneers [17] and Humane Intelligence [20], the U.S. Department of Defense (DoD) [18], and a crowdsourced cybersecurity platform, Bugcrowd [19]. All bounty programmes were introduced between 2021 and 2024. We describe the overall characteristics of these bias bounties below.

**Format.** All programmes were launched as time-bound contests, with Twitter [16] and the Bias Buccaneers [17] each hosting a single competition; Humane Intelligence [20] hosting multiple challenges; the DoD [18] announcing in 2024 that there would be two bias bounties; and Bugcrowd [21] offering to run competition-style bias assessments for its customer organisations.

**Participation.** The public participation model applies to four bounties [16-18, 20], while Bugcrowd's programme is private and open for 'trusted, 3rd-party security researchers' [19]. Some programmes imposed eligibility criteria: the DoD limited participation to U.S.-based contributors [22]; Humane Intelligence required participants to be at least 18 years old and to sign a waiver for appropriate data use in their second challenge [23]; and challenges had formal requirements, such as requiring a HackerOne account to participate in Twitter's contest [24].

**Programme management.** Two of the bias bounty initiatives were identified as 'self-managed' (i.e., delivered in-house) [17, 20], while two others were classified as 'mixed management' [16, 18], as organisations partnered with crowdsourced cybersecurity platform providers. Bugcrowd's programme offers a platform for reward-based AI bias assessments and was categorised as 'platform-managed' [19].

**Access.** The access was 'open box' (i.e., 'Additional access, either organizational or technical, is provided to enable a deeper level of testing.') [15] in Twitter's programme [16]. In the DoD's challenge [18], participants testing for bias in an open-source large language model (LLM) had 'closed box' access (i.e., 'Testing is limited to publicly available resources and tooling, without additional organizational or technical access (e.g., to documentation or source code).') [15]. In Bugcrowd's initiative, access is dependent on the customer [19]. Humane Intelligence provided access to datasets (not models) for each of the three challenges, which were either publicly available [25], required signing a waiver [23], or required signing up for the competition [26]. The Bias Buccaneers prepared a dataset specifically for the bias bounty challenge, which is not available beyond that [27].

**Disclosure model.** Twitter [24] and the Bias Buccaneers [17] allowed delayed full disclosure of participants' findings (i.e., 'Researchers can freely disclose their findings to the public on a predetermined time frame without additional approval from the affected organization.') [15]. The disclosure model in the other three programmes could not definitively be determined based on publicly available information.

**Contest period.** Twitter's programme had the shortest period for working on and submitting bias reports, which was one week [16]. The contest period in Bugcrowd's initiative was considered customer-dependent, since it is a private offering to different client organisations [21], with contest periods that may vary in length. The other organisations set the period from one and a half to two months [17-18, 23, 25-26].

**Prize pool.** The total rewards ranged from $7,000 for five winners in Twitter's programme [16] to $24,000 for three winners and those that passed the qualification stage in DOD's programme [22].

Each programme is described in the following subsections.

Table 1: Characteristics of the identified AI bias bounty programmes

| Year | Format | Participants | Programme Management | Access | Disclosure Model | Contest Period | Prize Pool |
|---|---|---|---|---|---|---|---|
| 2021 [16, 24] | Time-limited contest | 'anyone with a HackerOne account' | Mixed management | 'Open box' | Delayed full disclosure | 1 week | $7,000 in total |
| 2022 [17, 27] | Time-limited contest | No eligibility criteria known | Self-managed | Dataset created for the contest | Delayed full disclosure | 1,5 months | $16,000 in total |

| Year | Format | Participants | Programme Management | Access | Disclosure Model | Contest Period | Prize Pool |
|---|---|---|---|---|---|---|---|
| 2024 [18, 22] | Time-limited contest | 'all [U.S.-based] members of the public' | Mixed management | 'Closed box' | Unclear | 1,5 months | $24,000 in total |
| 2024 [19, 21] | Time-limited contests | 'trusted, 3rd-party security researchers (aka a "crowd") with specialized tools and skills in prompt engineering' | Platform-managed | Customer-dependent | Unclear | Customer-dependent | Depends on bias |
| 2024 [20, 23, 25-26] | Time-limited contests | Contests 1 & 3: no stated eligibility criteria; contest 2: age 18+, signed waiver for appropriate data use | Self-managed | Datasets provided, access challenge-dependent | Unclear | 1st and 3rd contest: 2 months; 2nd contest: 1,5 months | $9,900 - $10,000 per contest |

### 3.1.1. Twitter

In 2020, many users reported concerns about racial bias in Twitter's machine learning (ML)-based image cropping tool [28]. Twitter (now X) subsequently performed an internal bias analysis [29] and partially decommissioned the tool [30]. Following those events, Twitter introduced an algorithmic bias bounty challenge in 2021 [16]. To identify algorithmic harms, participants received access to their saliency model and code used to crop an image for display in previews based on the predicted most salient point [24]. They were encouraged to use Twitter's paper presenting their internal analysis [29] and the associated code as reference to assess biases in the image cropping algorithm while making 'a substantial novel contribution' beyond the findings in their paper [24]. Submissions included two components: 1) a read-me file detailing the identified harm, the importance of the harm, the findings supported by a description of the qualitative and quantitative methods, and a self-grading recommendation using the rubric; and 2) a GitHub link with code and necessary data or image files to reproduce and verify the harm. Participants' submissions were graded based on the type of harm, differentiating between unintentional harms (i.e., biases, discrimination, or other related harms from natural images) and intentional harms (i.e., arising from doctored images) [24]. Unintentional harms were awarded more points, with denigration, stereotyping, and underrepresentation scoring more in both harm categories (i.e., intentional and unintentional) compared to reputational, psychological, and economical harms. Several grading criteria (i.e., harm score multipliers) were used by four judges. Higher scores were given across the following criteria: harm measurement across multiple identity axes and a higher severity of the harm's damage or impact; a larger number of affected users; a higher likelihood of the harm (only applied to unintentional harms); lower effort and skill requirements to create an attack (i.e., exploitability; only applied to intentional harms); a strong justification of the harm's importance with a well-motivated methodology; a conclusive demonstration of the risk of the harm (i.e., clarity of contribution); and creativity. Participants submitted their contributions through the security platform HackerOne, with whom Twitter collaborated. They had one week to work on their submission. The prize pool included $3,500 for first place, $1,000 each for second place and for the single 'Most Innovative' and 'Most Generalizable' submissions, and $500 for third place [24].

The winners of the competition were a then-PhD student in computer science; the company HALT AI; an individual who is an electrical engineer and technology and human rights researcher; an ML engineer; and an anonymous participant [31]. In a blog post, two of the challenge organisers reflected on the learnings, highlighting submissions that addressed a variety of ML biases beyond race and gender. They noted contributions from individuals and groups worldwide, including universities, startups, enterprises, and participants with diverse backgrounds, some of whom did not have ML expertise [32]. Twitter (or X) has not organised further public bias bounties.

As the first publicly known AI bias bounty, the programme drew multiple comments. The Algorithmic Justice League's report on bug bounties for algorithmic harms acknowledged its 'potentially wide-reaching value' but highlighted limitations, including the lack of prior work cited on adapting bug bounty best practices and no recognition or reward for individuals who raised initial concerns before the programme. Additionally, the decision of not publishing the calculated scores for the winning submissions was seen as 'a missed opportunity for transparency around how their scoring framework was applied in practice, and likely constitutes a barrier to their framework being adopted and built upon by others' [15]. The authors also suggested adding clear definitions for relevant terms in the scoring framework and providing stronger justification for the score weightings of harms. For instance, rewarding a larger number of affected people in the scoring system disincentivises the search for biases affecting smaller groups (e.g., trans people), potentially leading to their specific harms being overlooked. The algorithm was argued to be likely low-risk to test, as harms had been exposed on social media, internal research had found flaws and been published, and the algorithm had been partially decommissioned. Also, the system being open-source meant no disclosure of proprietary information was required [15].

Kayser-Bril of AlgorithmWatch, a nonprofit organisation advocating for responsible algorithms and AI, stated, 'Although Twitter should be lauded for this experiment, it remains far short of addressing the harms caused by algorithmic bias.' The position argued that the challenge focused on a 'relatively unimportant algorithm' already accessible to users, while more impactful algorithms, like those driving the timeline or ad targeting, remain closed [33]. Moreover, the programme offered a modest $7,000 prize pool for five winners, leaving about 25 participants unrewarded. Unlike security bug bounties, which offer higher payouts due to clear regulatory and financial incentives, algorithmic bias lacks similar motivations, as those impacted do not constitute paying customers and regulatory enforcement remains weak [33]. Twitter users commenting on the challenge announcement also regarded the prize money as low [34].

A conference article by Lopez also addressed Twitter's bias bounty, highlighting limitations and power dynamics. Participants were granted limited access to the saliency-based prediction model. They could input images and receive saliency heat maps, their maximum saliency scores, and cropping outputs, but they were not given transparent insight into the tool's inner workings. Lopez stated:

> 'The imbalance of resources and knowledge between Twitter as a company on the one hand, and its users and the bias bounty participants on the other hand, makes it possible for Twitter to claim transparency, yet not grant meaningful access to the inner workings of its saliency based cropping tool.' [35]

The article also noted that Twitter benefited disproportionately by crowdsourcing knowledge production from participants, many of whom went uncompensated. Ultimately, the programme served as a form of 'counter-resistance', stabilising Twitter's position within the bias discourse [35].

### 3.1.2. Bias Buccaneers

In 2022, the nonprofit organisation Bias Buccaneers (whose founders had previously organised Twitter's bias bounty programme) introduced a challenge for participants to 'build a machine learning (ML) model that labels each image [of synthetically generated human faces] with its most likely skin tone, perceived gender, and age group' [17]. Participants were provided with a dataset of 15,000 images, including partially labelled training data and a separate test set of only labelled images, and could use a supervised, an unsupervised, or a hybrid approach to train their models (i.e., using labelled images, unlabelled images, or a mix of both). The scoring rubric was based on four grading criteria: model accuracy for the demographic attributes, classification accuracy disparity across classes for each demographic attribute, randomness in tags for noisy samples, and efficiency of the code in producing predictions. The contest period was set for about one and a half months, with rewards of $6,000 for first place, $4,000 for second place, $2,000 for third place, and $4,000 for the best unsupervised model. Regarding the rights for the submission's future use, it was stated: 'You retain full rights to your submission, including the choice of license, a public or private repository, and rights to its future use.' [17]

We could not find publicly available information on the outcomes of the challenge, including the announcement of the winners. Additionally, we found no other challenges by the organisation, and their website was inaccessible when reviewing this bias bounty.

### 3.1.3. U.S. Department of Defense

Another AI bias bounty was organised by the DoD's Chief Digital and Artificial Intelligence Office's Responsible AI division in collaboration with companies ConductorAI and Bugcrowd. The programme focused on identifying biases in an LLM by evaluating its responses within a chat interface, aiming to uncover preferential or discriminatory outputs related to protected classes (e.g., based on age, gender, or sexual orientation) pertinent to the DoD [36]. It was open to all members of the public in the U.S. and ran for one and a half months [22]. According to the DoD, no coding experience was required to join the contest [18]. It offered a prize pool of $24,000: $9,000 for first place, $5,000 for second, $2,000 for third, and $250 for participants who progressed through the qualification stage [22]. Keith Hoodlet, a security professional who won first place, identified biases in the LLM that included discrimination against pregnant women and a tendency to assume superior officers were male, frequently answering with 'yes, sir!' [36]. While a bounty brief with information on the submission evaluation rubric, qualification criteria, and terms of the challenge was published [22], we were unable to access it. The DoD stated that this bias bounty was the first of two [18], though we could not find any information on the second programme at the time of this survey.

### 3.1.4. Bugcrowd

In 2024, Bugcrowd launched an AI bias assessment service, which follows the form of a bounty programme due to its reward-based approach for third-party individuals to identify biases in LLMs, such as representation or algorithmic bias. The company describes their AI bias assessment service as 'a private, time-bound engagement with duration, scope, severity scoring, and competition-style rewards structure determined by you with guidance from Bugcrowd' [21]. Using their platform, third-party security researchers skilled in prompt engineering identify and report biases in both open-source and private AI models for companies from all industries and government agencies. Findings are validated and ranked using Bugcrowd's bias severity rating system, with researchers compensated based on the validated impact of their findings; more severe biases receive higher payouts [19].

### 3.1.5. Humane Intelligence

The nonprofit organisation Humane Intelligence, led by Rumman Chowdhury, whose team introduced Twitter's bias bounty, has also organised bias bounty challenges. Participation is open globally to individuals and teams, with the option to find partners through their Discord channel [20].

Their first of ten planned challenges, launched in May 2024 and based on the evaluation and data from their previous Generative AI Red Teaming Challenge, aimed to involve bounty hunters in the detection of problematic LLM outputs. Participants could enter one of nine contests, corresponding to 'Beginner', 'Intermediate', and 'Advanced-level' competitions in each of the three categories: factuality (i.e., 'model's ability to discern reality from fiction and provide accurate outcomes', with a focus on harmful cases such as political or economic misinformation), bias (i.e., 'scenarios that would broadly be considered defamatory or socially unacceptable by perpetuating harmful stereotypes'), and misdirection (i.e., 'incorrect outputs and hallucinations that could misdirect or mislead the user') [25]. At the beginner level, participants identified gaps in a dataset (factuality, bias, or misdirection), proposed new categories of data to improve its representativeness, and created five prompts per data category that produce harmful outputs. Grading by Humane Intelligence staff was based on the number of violations generated by the prompts and the prompts' diversity. Winners in each category received $800. Building on the beginner level, participants at the intermediate level generated synthetic data to fill gaps in the dataset rather than manually creating prompts. Grading was based on the percentage of successful violations generated by the synthetic data and the diversity of proposed improvements, with winners receiving $1,000 per category. At the advanced level, participants developed a likelihood estimator to predict the probability of a prompt leading to a violation. As grading criteria, the performance (i.e., accuracy and F-score) of the model was evaluated using a holdout dataset, with $1,500 prizes available per category. Participants had two months to complete the challenge. The winners listed on the organisation's website [25] had backgrounds in AI engineering, mechanical engineering, data science, informatics, mathematics, and technology policy. Their backgrounds were identified by searching their LinkedIn profiles.

The second challenge, in partnership with the counterterrorism company Revontulet, focused on computer vision to detect and interpret hateful image-based propaganda, particularly from far-right extremist groups. The description noted, 'Due to the sensitive nature of this topic, we are only hosting intermediate and advanced challenges to skew towards more experience.' Intermediate-level participants were tasked with building an unsupervised ML model that identifies if an unlabelled image has extremist content. Participants at the advanced level built on the intermediate task by generating adversarial examples from the test dataset to test their model's robustness, exploring methods for adversarial example generation to trick the model, and making predictions on the perturbed images with their model. Submissions were scored based on accuracy against a holdout labelled dataset, accounting for flipped binary labels. The prize pool totalled $4,000 for three intermediate-level winners and $6,000 for three advanced-level winners [23]. Winners came from quantitative backgrounds, including AI engineering, security engineering, mechanical engineering, quantum computing, and business analytics.

The third bias bounty challenge was organised in collaboration with the Indian Forest Service and offered three levels of participation. The 'Thought Leadership Level' required participants to write a position piece exploring bias and fairness in tree planting site recommendations, proposing mitigation strategies, covering contestability, and comparing existing methods in the field. Participants entering the beginner-level competition were tasked with identifying the key features that affect tree planting feasibility in a given dataset, while those at the intermediate level were to build on these features to develop a site recommendation engine predicting the suitability of sites for tree planting. The prize pool consisted of $4,000 for three intermediate-level winners and $3,000 each for three beginner-level and three thought leadership-level winners [26]. Seven winners were announced [26], with backgrounds in ecology and sustainable development, computer science, robotics and cognitive systems, AI and ML, engineering, and economics and business analytics.

## 3.2. Academic Literature on AI Bias Bounties

Bias bounties, along with safety bounties, were included as one of the institutional mechanisms for supporting the verifiability of claims in trustworthy AI development [10]. The six issues outlined for establishing a bounty programme for AI systems include determining appropriate compensation rates that correspond to the severity of identified issues, establishing processes for soliciting and assessing submissions, ensuring timely disclosure of detected issues, designing appropriate interfaces for bias reporting, establishing processes for addressing and fixing reported issues, and preventing perverse incentives. The authors note that bounty programmes could increase scrutiny of AI but cannot ensure the trustworthiness of an AI system on their own and may not help to discover certain issues. Different vulnerability types and associated bounties may require different approaches depending on factors like the stakes involved and the potential for bias remediation. Bounties could also compensate for valuable information without immediate fixes available [10].

Another work explored queer perspectives on AI bias bounties through a participatory workshop that identified four themes [9]. The first theme, 'Queer Harms', highlights how bias bounties can harm queer individuals through a lack of attention to their concerns (as queer users are a minority) and participation risks, namely privacy violations (incl. forced outing) and exposure to harmful content. 'Control' as the second theme reflects concerns about the imbalance of power, with misaligned values between companies and users, disproportionate resource allocation, and the inability to specify nuanced levels of social context and cultural norms (incl. the challenge of encouraging participation from users from diverse backgrounds). The third theme, 'Accountability', describes companies running their own bias bounties as a potential conflict of interest and additionally notes the lack of incentives for companies to use the feedback from bias bounties. It calls for community ownership of bias bounties to focus on the needs of queer people and for external incentives that encourage companies to devote time and effort to implement ethical AI practices, including bias bounties. The theme of 'Limitations of Bias Bounties' critiques their effectiveness in mitigating biases and their accessibility. It highlights the limited ability to address the underlying causes of harms and to question the legitimacy of certain AI systems, the influence of subjective perspectives and political motivations on the view of harm severity, and the barriers to participation for non-technical communities, while recognising their value in decentralising bias detection. Thus, the article advocates for community-driven bias auditing, emphasising the need for ownership, co-design, and the redistribution of power to marginalised groups. In fact, the authors argue 'that unless power disparities between companies and marginalized communities are minimized (i.e., communities own bias bounties), bounties cannot be effective' [9].

Globus-Harris et al. published an algorithmic framework for conducting bias bounties for ML models [37]. Under this framework, 'bias hunters' identify subgroups where an ML model performs suboptimally and propose improved models for those subgroups. The combination of subgroup and improved model is referred to as 'a certificate of sub-optimality'. The framework provides an algorithm to create an augmented model from the submitted certificate that reduces both subgroup-specific and overall errors without trade-offs. Unlike other fair ML frameworks with fixed constraints, this approach allows for the dynamic definition of flexible and arbitrarily complex certificates. It leverages adaptive data analysis techniques (e.g., description-length methods) to validate submitted fixes on a held-out test set to mitigate the risk of overfitting. The algorithmic framework guarantees convergence to the Bayes optimal model and provides techniques to verify the correctness of submitted certificates. Bounties can be determined based on the size of the subgroup identified and the magnitude of the performance improvement achieved, with total payouts designed to be bounded in advance. The framework's primary limitation is its reliance on existing data distributions, meaning it cannot address biases from unrepresentative sampling or missing key predictive features, though it can identify areas needing new data collection. A preliminary deployment with 83 students in 36 teams found that hybrid approaches combining manual data analytics and automated methods outperformed purely algorithmic approaches, highlighting the value of engaging a community to uncover issues that automated methods alone might miss [37].

In addition to these works, the Algorithmic Justice League's report on bug bounties for algorithmic harms [15] and the article by Lopez addressing Twitter's bias bounty [35], we found multiple academic publications that cite existing bias bounty efforts in the context of third-party algorithmic auditing [e.g., 38-42].

## 4. Discussion

This survey aimed at providing an overview of the current state of AI bias bounties and found five bounty programmes, five research publications, and further articles citing these existing efforts. Thus, bias bounties are still a nascent AI practice.

### 4.1. Challenges of AI Bias Bounties and Ways Forward

While four of the five bias bounties organised as time-limited challenges were open to the public, including those without coding experience, the winners primarily had quantitative backgrounds, frequently related to AI. Though it is plausible that having AI-related experience increases the likelihood of winning such contests, and bias bounties are still in their infancy, future work should explore how underrepresented groups, particularly those affected by AI bias and with little to no coding experience, perceive the 'unwritten' technical entry requirements and their chances of winning. Moreover, finding ways to lower the entry barriers for those without an AI background is crucial. We see Humane Intelligence's approach of offering different competition categories with different requirements [25-26] as a step in this direction. Further, we propose that organisers should complement their bounties with measures to improve AI literacy to reach those particularly affected by bias and prevent the reinforcement of the existing dominance of white, cisgender men in the AI field [43-45]. For instance, Humane Intelligence has a webpage dedicated to resources, including reports, guides, and tools related to AI and bias bounties, as well as video tutorials for beginners covering dataset downloading, coding notebook creation, data analysis, and the process of submitting findings [46].

We also found that two organisations managed the delivery of their programmes and the handling of submissions themselves [17, 20], two others partnered with a crowdsourced cybersecurity platform [16, 18], and another initiative connecting organisations and security researchers was organised directly by one of these platforms [19]. Third-party platforms as intermediaries can provide expertise, oversight and conflict mediation, and logistical support for bias bounties (e.g., submission management, reward processing) but also risk suppressing findings and prioritising corporate interests, as they rely on these organisations as their customers [15]. This is one example of where bias bounties may learn from the bug bounty model, though differences exist in the financial and regulatory incentives to establish a bounty programme [33]. Further work should build on the Algorithmic Justice League's report on adapting bug bounties to algorithmic harms [15] and explore how a fair mediation between organisations owning an AI and bounty hunters can be achieved through third-party platforms. Efforts also exist in related approaches, such as AI safety bounties [10], and drawing lessons from these approaches remains important. Moreover, sharing decisions and learnings regarding the design and implementation of a bias bounty would help others adapt these efforts and identify areas for improvement [15].

Contrary to the prediction made by the market research firm Forrester that many major companies would offer bias bounties or similar programmes in 2022 [47], Twitter remains the only company to have organised a bias bounty for its own AI tool in 2021. AI scientist Lance Eliot described potential bias bounty adoption barriers for organisations, including concerns that they imply an admission of bias, risk public exposure of flaws and reputational damage (we note that media coverage may focus more on the biases identified rather than the initiative to organise a bounty programme), or enable malicious activities like cyberattacks. He also highlighted scenarios where bounty hunters exploit findings for personal gain, such as by pressuring a company for larger payments [48]. The risk of being overwhelmed by the amount of submissions and false reports, the labour-intensive process of validating claims [11, 48], and defining AI biases precisely enough to provide clarity without being so restrictive that potential biases are overlooked are further challenges. Moreover, the reward needs to be high enough to attract participants without incentivising malicious behaviour, though very large rewards can lead to an increase in false positive reports due to superficial investigations by a higher number of bounty hunters. Furthermore, organisations must find a balance in ensuring accessibility to the model for bounty hunters, as granting sufficient access to identify biases must be weighed against maintaining cybersecurity protections and may involve restrictive legal requirements that could deter potential participants [48]. Finally, a bias bounty programme could uncover the need for redevelopment (e.g., starting with an entirely new dataset) [48] or decommissioning of the AI model, potentially conflicting with companies' economic interests.

Given the promise of first programmes, despite shortcomings highlighted in Twitter's challenge, we argue for increased adoption of bias bounties, while noting that addressing barriers for organisations could have negative consequences for those affected by AI harms and for potential bounty hunters. For instance, a company might implement stricter entry requirements to prevent an influx of 'low-quality' or inaccurate submissions, which could exclude those without coding experience but with potentially underrepresented perspectives. Thus, a bias bounty might be a 'win-win' experience for the organisation and participants but could come at the expense of reinforcing existing power imbalances in the AI field. Moreover, a company might request a non-disclosure agreement in a programme to protect intellectual property and sensitive data, prevent premature disclosure of bias, and avoid misuse of access. This could conflict with a philosophical opposition to secrecy, concerns about the company's lack of public accountability for biases, a perceived power imbalance where the company controls the information and narrative, and discomfort with legal obligations. It may be argued that bias bounty co-design, proposed in one of the presented articles [9] as a potential measure for redistributing power to marginalised groups, is unfeasible at this stage since it might be perceived as an additional burden by adopters of such programmes. However, we suggest that case studies of bias bounty co-design with AI users, including underrepresented groups, could identify ways to address the conflicting interests between bias bounty organisers and users. Such insights would facilitate bias bounty adoption. To encourage a redistribution of power between companies and users, bias bounty co-design projects should go beyond tokenism, as illustrated in Arnstein's ladder of citizen participation [49].

### 4.2. Strengths and Limitations

A strength of this survey is the presentation of the current state of AI bias bounties, which is a novel and underexplored area. However, while this survey descriptively presents existing efforts, it does not evaluate the quality of AI bias bounty programmes. Including commentaries on Twitter's bias bounty helped identify its limitations but does not replace a systematic quality assessment of all identified AI bias bounty programmes.

Another limitation is the exclusive search for material in English, which may have resulted in missing AI bias bounty programmes and documents. Consequently, the finding that all identified organisations conducting bias bounties are U.S.-based may also be influenced by this search strategy. Moreover, the searches in PhilPapers and IEEE Xplore did not yield any results. Thus, only results from Google and Google Scholar were screened. We believe that the search strings used for PhilPapers and IEEE Xplore were broad, and the absence of hits likely reflects the modest number of publications on bias bounties. Also, all 216 results returned by Google were screened, thereby minimising the risk of missing relevant material.

The bias bounty programmes and research publications were included and reviewed by a single researcher (SK). However, wrongful exclusion of bias bounties or research is unlikely due to their overall scarcity and the clear definition of what constitutes a bias bounty. Moreover, while the review of bias bounties and research publications by a single researcher is a limitation, the risk of a biased presentation of the state of AI bias bounties is reasonably low, as the aim was to provide a descriptive overview of existing efforts, which is less prone to errors due to misinterpretation than a quality appraisal of efforts. Parts of the overview, such as the summaries of [9] and [37], were additionally validated by another researcher (ND) to ensure an accurate presentation of previous works.

## 4.3. Recommendations

In summary, we recommend the following actions for bias bounty organisers and researchers to advance AI bias bounties as a method to identify and mitigate biases:

- evaluate the quality of the existing AI bias bounty programmes to inform the design of future bias bounties
- explore ways to lower the entry barriers for people who are (1) underrepresented in the AI field, (2) without a coding background, and/or (3) affected by AI bias; this should include investigating how they perceive (a) the requirements for entering a bias bounty competition and (b) their chances of winning, as well as undertaking measures to improve AI literacy (e.g., educational videos) when carrying out a bias bounty programme that is open to the public
- build on existing work that adapted bug bounties to algorithmic harms [15] and investigate how a fair mediation between AI-owning organisations and bounty hunters can be achieved through third-party platforms
- share decisions and learnings related to the organisation and outcomes of a bias bounty, including but not limited to:
  - bias bounty design foundations, such as references and frameworks used
  - rules for public disclosure of findings (incl. when and how participants can share their findings)
  - clear definitions and rationales behind the scoring system for submissions
  - calculated scores for winning submissions, illustrating how the scoring system was applied
  - implementation takeaways, such as observed successes and challenges
- explore how barriers to AI bias bounty adoption can be lowered for organisations, while ensuring that this does not negatively affect bias bounty hunters and those impacted by AI bias; this should include conducting case studies of bias bounty co-design to identify how conflicting interests between AI-owning organisations and bounty hunters can be addressed

## 5. Conclusion

AI systems, developed by humans who may inadvertently introduce biases and trained on real-world data that can reflect societal biases, require external scrutiny to ensure fairness. Rewarding individuals or organisations for identifying bias through bounty programmes could be one way to increase the scrutiny of AI. With five bias bounty programmes organised as time-limited contests and five research publications contributing to this area, the adoption of bias bounties and the related scientific discourse remain modest. Future efforts should further explore the transferability of best practices from bug bounties to bias bounties. They should also examine how such programmes can be designed to be sensitive to underrepresented groups while addressing conflicting interests with potential AI bias bounty organisers, thereby fostering their responsible adoption.

**Acknowledgements**
The authors' contributions were as follows: Conceptualisation: SK, FH. Methodology: SK. Formal analysis and validation: SK, ND. Writing – original draft: SK. Writing – review & editing: SK, ND, FH. Supervision: FH. All authors read and approved the final version. We would like to thank the authors of the Algorithmic Justice League's report on bug bounties for algorithmic harms for kindly granting permission to use their figure showing the bug bounty programme design levers, which were used in our survey for analysing existing bias bounty programmes. Moreover, we thank Keith Hoodlet, winner of the DoD's AI bias bounty contest, for helping us verify our understanding of participants' access for bias testing (i.e., 'closed box' vs. 'open box') in the DoD's contest.

# A    Appendix: Search Strategy and Results

| Databases/ Search Engine | Query String | Search Results Screened |
| --- | --- | --- |
| Google | ("artificial intelligence" OR "AI" OR "machine learning" OR "ML" OR "algorithm*") AND ("bias") AND ("bounty" or "bounties") | 216 |
| Google Scholar | ("artificial intelligence" OR "AI" OR "machine learning" OR "ML" OR "algorithm*") AND ("bias") AND ("bounty" or "bounties") | 100 from 1,200 results (i.e., stopped screening results after page 10, as relevance of search results decreased considerably) |
| IEEE Xplore | ("Abstract":"artificial intelligence" OR "Abstract":"AI" OR "Abstract":"machine learning" OR "Abstract":"ML" OR "Abstract":"algorithm*") AND ("Abstract":"bias") AND ("Abstract":"bounty" OR "Abstract":"bounties") | None |
| PhilPapers | (bias) AND (bounty OR bounties) in Philosophy of Cognitive Science > Philosophy of Artificial Intelligence | None |